\documentclass[sigconf,authorversion,nonacm]{acmart}

\usepackage{algorithmic}
\usepackage{graphicx}
\usepackage{textcomp}
\usepackage{xcolor}
\usepackage{listings}
\usepackage{hyperref}
\usepackage{makecell}
\usepackage{multirow}
\usepackage{tikz}

\setcopyright{none}

\begin{document}

\title{Exploring the Versal AI engines for accelerating stencil-based atmospheric advection simulation}


\author{Nick Brown}
\email{n.brown@epcc.ed.ac.uk}
\orcid{0000-0003-2925-7275}
\affiliation{%
  \institution{EPCC at the University of Edinburgh}
  \streetaddress{47 Potterrow}
  \city{Edinburgh}
  \country{UK}
}


\begin{abstract}
AMD Xilinx's new Versal Adaptive Compute Acceleration Platform (ACAP) is an FPGA architecture combining reconfigurable fabric with other on-chip hardened compute resources. AI engines are one of these and, by operating in a highly vectorized manner, they provide significant raw compute that is potentially beneficial for a range of workloads including HPC simulation. However, this technology is still early-on, and as yet unproven for accelerating HPC codes, with a lack of benchmarking and best practice.

This paper presents an experience report, exploring porting of the Piacsek and Williams (PW) advection scheme onto the Versal ACAP, using the chip's AI engines to accelerate the compute. A stencil-based algorithm, advection is commonplace in atmospheric modelling, including several Met Office codes who initially developed this scheme. Using this algorithm as a vehicle, we explore optimal approaches for structuring AI engine compute kernels and how best to interface the AI engines with programmable logic. Evaluating performance using a VCK5000 against non-AI engine FPGA configurations on the VCK5000 and Alveo U280, as well as a 24-core Xeon Platinum Cascade Lake CPU and Nvidia V100 GPU, we found that whilst the number of channels between the fabric and AI engines are a limitation, by leveraging the ACAP we can double performance compared to an Alveo U280.
\end{abstract}


\maketitle

\section{Introduction}
The Versal Adaptive Compute Acceleration Platform (ACAP) is a new type of FPGA which combines Programmable Logic (PL) with other facets including CPU-based Programmable Subsystem (PS) and AI engines \cite{gaide2019xilinx}. These AI Engines, or AIEs and we use these two terms interchangeably throughout this paper, are of specific interest here as they are designed to accelerate highly-parallel vector operations. The Versal AI-series contains up to 400 engines running between 1 and 1.2 GHz, and each engine follows a Very Long Instruction Word (VLIW) design, capable of issuing seven instructions per cycle. AI engines are capable of undertaking 8-way vectorized single-precision floating point operations and up to 128 8 bit fixed point arithmetic operations per cycle.

The large amount of raw compute provided by the AIEs is interesting for High Performance Computing (HPC) workloads, where the ability to use the Versal's PL to tailor memory accesses bespoke to an application and the AI engines to accelerate the compute has potential. To date there have been a very limited number of preliminary AIE studies \cite{lee2021preliminary} \cite{zhang2022h}, and-so an important outstanding question is whether these engines can be effectively leveraged for real world HPC kernels. In this work we use the atmospheric advection kernel of the Met Office NERC Cloud model (MONC) \cite{easc}, which is an open source high resolution atmospheric modelling framework, as a vehicle to explore the AI engines. Following a stencil-based compute pattern, which is very common in HPC codes, in this short paper we explore how to best map this compute pattern onto the AIEs and how performance compares against other approaches. This paper is structured as follows, in Section \ref{sec:bg} we explore the background to this work before summarising the experimental setup in Section \ref{sec:experimental_setup}. Section \ref{sec:aie_focus} explores structuring our AIE kernel(s) and interfacing these with the PL, before undertaking a performance comparison against other hardware in Section \ref{sec:multiple}. We then conclude and discuss recommendations in Section \ref{sec:conclusions}.

The novel contributions of this paper are \textbf{1)} An exploration of techniques to most effectively structure AIE kernels \textbf{2)} An initial performance comparison between the AIEs and other hardware \textbf{3)} Highlighting some of the limitations of the current AIE technology that one must consider when working with the hardware. 

\section{Background}
\label{sec:bg}
\subsection{The Versal AI engines}
\label{sec:aie}
The VLIW design of Xilinx's new AI engines is such that, per cycle, each engine is capable of issuing a maximum of two loads, one store, one scalar operation, one fixed point or floating point vector operation, and two move instructions. The vector unit is of size 256 bits, and focusing on single precision floating point arithmetic in this paper, each engine is capable of undertaking up to eight single precision floating point calculations per cycle. Consequently it is important to ensure code is correctly vectorized to obtain best performance on the AIEs. Based on 400 AI engines running at 1.2GHz on the VCK5000, there is a theoretical single precision floating point performance of 3.6 TFLOPS.

AI engines are arranged in a 2D array, with engines connected to their neighbours in both dimensions. Each engine contains 16KB of program memory and 32KB of local data memory and, for the later, is able to directly access the memories of three of its neighbours providing a total of 128KB contiguous addressable data memory \cite{aie_best_practice}. Furthermore, each engine has two 32 bit input streams and two 32 bit output streams which are combined with a FIFO to provide 128 bit access every four clock cycles. Lastly, AI engines connect to one of their neighbours via a cascade stream which is 384 bits wide and designed to allow arithmetic operations to be chained.

AIE code comprises two parts, kernels which will be mapped to AI engines and a graph description which connects kernels together via their streams and memories, as well as to the PL. Programmatically there are two ways in which data can enter or leave a kernel, windows and streams \cite{aie_programming_guide}. Windows provide a buffer, where the current data position in the window is tracked. For input windows data is consumed from this buffer by the kernel, for output windows data is written. The other approach, a stream, provides an infinite number of scalars and vectors that can be read and written by the kernel. There underlies an important difference between these two approaches, where a window of data will only progress to the next window between outer iterations of the kernel, as driven by the AIE graph, whereas streams can continually be read from and written to inside the kernel. Consequently, with windows one must frequently start and stop their kernels to refresh the window data, which is not required with streams.

Whilst the AI engines are the major focus in this paper, it is also important to highlight the general architectural improvements that Xilinx have made to the PL in their Versal series. Built on a 7nm process technology, numerous components including DDR controllers and PCIe interface have been hardened compared to previous generations \cite{ahmad2019xilinx}. Furthermore, a dedicated Network on Chip (NoC) is provided which not only connects the PL with the AIEs, but can also be used between IP blocks on the PL. 

\subsection{Piacsek and Williams advection kernel}

Advection is the movement of values through the atmosphere due to wind and, at around 40\% of the runtime, is the single longest running piece of functionality in the MONC model \cite{easc}. The code loops over three fields;  \emph{U}, \emph{V} and \emph{W}, representing wind velocity in the \emph{x}, \emph{y} and \emph{z} dimensions respectively. This Piacsek and Williams (PW) \cite{pwadvection} advection scheme is called each timestep of the model and calculates advection results, otherwise known as \emph{source terms}, for each field. This advection scheme is a stencil based algorithm, of depth one, where calculating the value of a grid cell requires contributions from neighbouring values across all three dimensions.

\begin{figure}[h]
  \centering
  \includegraphics[scale=0.55]{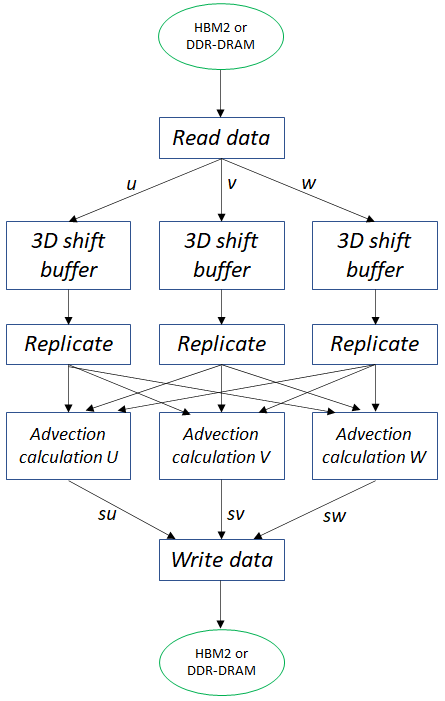}
  \caption{Dataflow design of HLS advection kernel from \cite{brown2021accelerating}}
  \label{fig:df_design}
\end{figure}

In previous work \cite{brown2021accelerating} this kernel was ported to an Alveo U280 using High Level Synthesis (HLS) and leveraging the \emph{dataflow} HLS pragma to run multiple components concurrently. The structure of this HLS kernel is illustrated in Figure \ref{fig:df_design}, where the boxes are dataflow regions and arrows between these are internal HLS streams. 3D shift buffers provide a bespoke memory solution which is capable of delivering all 27-stencil values per cycle to the advection compute stages, which was found to be the optimal approach even though not all 27 neighbouring stencil values are required by the advection calculations. Given this existing structure it was our hypothesis that we could replace the advection calculation stages with streams to and from the AI engines, still leveraging the existing tailoring of memory accesses on the PL that worked well in \cite{brown2021accelerating}, with the raw compute power of the AI engines.

\section{Experimental setup}
\label{sec:experimental_setup}
In this work we are using a Xilinx VCK5000 containing a Versal VC1902 ACAP and 16GB of DDR4-DRAM. All VCK5000 runs are built using Vitis 2022.1, the PL is running at 300MHz, and the VC1902 contains 400 AI engines running at 1.2GHz. We compare against an Alveo U280 which contains 8GB of HBM2, is also running at 300MHz, and Alveo kernels are built using Vitis 2021.1. Both the VCK5000 and Alveo U280 are PCIe based cards hosted by a machine containing a 32-core AMD EPYC 7502 processor and 256GB DRAM. 

All reported results are averaged over five runs and performance results are reported as useful FLOPS, which is the number of floating point operations undertaken that contribute to the calculation's result. Our performance numbers measure device-side execution time only and do not include the time taken to copy input data to, or result data from, the host and device. This is because we are most interested in the performance of the AIEs in this work, and device-side performance therefore provides a clearer picture when comparing against other technologies that exhibit different host-device data transfer overheads.

\section{AIE porting and optimisation}
\label{sec:aie_focus}
We started by decomposing the advection stencil-based calculation into constituent operations, resulting in, for each grid cell, the code undertaking six additions, followed by six multiplications, then four subtractions and finally an addition reduction to sum these subtractions together. A floating point vector of size six is not supported by the tooling and-so we pad with an additional two empty values to make a vector of size eight. This is why we report useful, rather than total, FLOPS, as useful FLOPS ignores the processing of these empty values by only considering those floating point operations that actually contribute to the advection result.

\begin{figure}[h]
  \centering
  \includegraphics[width=\linewidth]{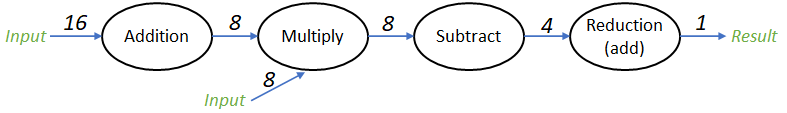}
  \caption{Illustration of AIE calculations per grid cell, with the numbers representing the number of single precision floating point numbers provided.}
  \label{fig:calc}
\end{figure}

The structure of this kernel is illustrated in Figure \ref{fig:calc}, with the first 8-way vector addition requiring sixteen floating point numbers comprising the operands. The multiplication requires an additional eight input numbers which are multiplied by the result of the preceding addition. We packaged this as a single AIE kernel and Listing \ref{lst:kernel_code} provides a partial sketch of the code. In order to prepare for the vector addition, streams of four numbers are read and loaded into the appropriate locations of the \emph{lhs} and \emph{rhs} vectors in lines 11 to 14. These vectors are then provided as arguments to the \emph{aie::add} method at line 16, which undertakes the vectorized addition. Multiplication, subtraction, and reductions operations are handled similarly and omitted for brevity. It can be seen at line 6 that we are looping over grid cells, and the directives at lines 7 and 8 instruct the AIE compiler to undertake software pipelining where possible, attempting to keep the VLIW slots filled as per Xilinx's best practice \cite{aie_best_practice}.

\begin{lstlisting}[frame=lines,caption={Sketch of AIE advection kernel code}, label={lst:kernel_code}, numbers=left]
void cell_advection(input_stream<float> * __restrict in_A, input_stream<float> * __restrict in_B, output_stream<float> * __restrict out) {
  aie::vector<float, 4> in_data;  
  in_data=readincr_v<4>(in_A);  
  
  int32 cells=(int32) in_data.get(0);  
  for (int i=0;i<cells;i++) 
  chess_prepare_for_pipelining
  chess_loop_range(64,) {
    aie::vector<float,8> lhs_nums, rhs_nums;  
    
    lhs_nums.insert(0,readincr_v<4>(in_A));
    lhs_nums.insert(1,readincr_v<4>(in_A));
    rhs_nums.insert(0,readincr_v<4>(in_B));
    rhs_nums.insert(1,readincr_v<4>(in_B));
    
    aie::vector<float,8> vadd=aie::add(lhs_nums,rhs_nums);
    ....
  }
\end{lstlisting}

The AIE API provides adaptive dataflow graphs which enables parameters to be dynamically set at runtime. However this is not supported by the VCK5000 shell and as such an alternative was required for setting the number of loop iterations at line 6. This is the reason that, for lines 2 to 5 in Listing \ref{lst:kernel_code}, four floating point numbers are read from the \emph{in\_A} stream and the first of these is extracted, cast to an integer, and used as the number of grid cells to loop over (this corresponding value has been streamed from the PL on start up). We must read the number of cells as a float because there are a maximum of two inputs and two outputs per kernel, and both inputs are required for the loading of operands. 

\begin{lstlisting}[frame=lines,caption={Sketch of AIE graph building code}, label={lst:graph_code}, numbers=left]
class simpleGraph : public graph {
private:
  kernel cell_advection_kernel[3];
public:
  input_plio in_A[3], in_B[3];
  output_plio out[3];

  simpleGraph(){
    cell_advection_kernel[0]=kernel::create(cell_advection);
    ...
    in_A[0] = input_plio::create("krnl_0_in0", plio_128_bits, "data/input_A.txt");
    in_B[0]  = input_plio::create("krnl_0_in1", plio_128_bits, "data/input_B.txt");
    out[0] = output_plio::create("krnl_0_out1", plio_32_bits, "data/output_0.txt");
    ...
    for (int i=0;i<3;i++) {
      connect<stream>(in_A[i].out[0], cell_advection_kernel[i].in[0]);
      connect<stream>(in_B[i].out[0], cell_advection_kernel[i].in[1]);
      connect<stream>(cell_advection_kernel[i].out[0], out[i].in[0]);
}}};
\end{lstlisting}

This code of Listing \ref{lst:graph_code} builds the high-level AIE graph, mapping kernels to individual AI engines. Three advection kernels are created, one for each field, and lines 11-13 defines the input and output ports between the PL and AIE for the first kernel (kernels two and three are omitted for brevity). These AIE ports are connected to the kernel ports in the loop at lines 15 to 19. As described in Section \ref{sec:aie}, physical connections between AI engines are 32 bits wide, but it can be seen for input ports we specify \emph{plio\_128\_bits} at lines 11 and 12. This directs the AIE compiler that streams on the PL are 128 bits wide (of type \emph{qdma\_axis<128,0,0,0>}) and therefore data will arrive in packets of 128 bits and be unpackaged into four 32 bit stream values. The reason for this is performance, where the PL is running much slower (in our case 300MHz) compared to the AIEs (1.2 GHz) and consequently in one clock cycle the PL is providing four 32 bit numbers which the AIE will then unpack per cycle. 128 bits is the maximum width supported, and this is why in Listing \ref{lst:kernel_code} the eight numbers comprising either side of the calculation are read via two \emph{readincr\_v} calls of size four at lines 11-12 and 13-14.

\begin{table}[htb]
 \begin{center}
  \begin{tabular}{|ccc|}
    \hline
    \textbf{Version} & \makecell{\textbf{Performance} \\ \textbf{\textit{(GFLOPS)}}} & \makecell{\textbf{Compared} \\ \textbf{to PL-only}} \\
    \hline
    PL-only (no AIEs) & 14.32  & -   \\
    \hline
    Initial        & 1.99  & 14\%   \\
    Multi-kernel & 4.06 & 28\% \\
    Cascade stream & 2.78 & 19\% \\
    Cascade multiplex & 3.87 & 27\% \\
    Multi-kernel windows & 0.91 & 6\% \\
    Chunking windows & 10.32 & 72\% \\
    Reduction on host & 16.13 & 113\% \\
    Double vectorization & 18.48 & 129\% \\
  \hline
\end{tabular}
  \caption{Compute performance of different versions of AIE design compared against PL-only non-AIE implementation. All runs undertaken in single precision floating point on Xilinx VCK5000 using a problem size of 67 million grid points.}
  \label{tab:aie_performance}
 \end{center}
\end{table}

It can be seen in Listing \ref{lst:graph_code} that there is a separate kernel instance created for each of the three fields, with each of these running on a separate AI engine. Whilst the calculations for each field are different, this difference lies in the specific stencil locations that are used, and the underlying arithmetic operations are the same. Consequently we are able to reuse the same kernel code, but provide different values to these from the PL side per field. The performance of this version is reported in Table \ref{tab:aie_performance} by the \emph{initial} row, and it can be seen that this was significantly slower compared to instead undertaking all arithmetic operations on the PL (\emph{PL-only (no AIEs)}).


\subsection{Optimising the data transfer}
The maximum 128 bit width of data between the AIEs and PL was a major reason for the poor performance of our initial version reported in Table \ref{tab:aie_performance}. This was because, per cycle, the PL was only able to stream four single precision floating point numbers per stream to the AIE, whereas 24 were required (16 for the addition and 8 for the multiplication). The number of inputs to an AIE kernel is limited to two, therefore meaning that the PL could provide a maximum of eight values per cycle. Consequently three writes on each stream were required per grid cell and this conflict resulted in an initiation interval of three in our HLS code on the PL.

\begin{figure}[h]
  \centering
  \includegraphics[width=\linewidth]{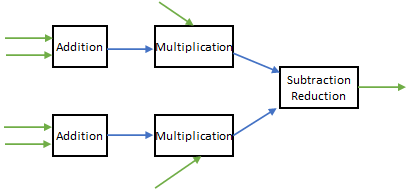}
  \caption{Multi-kernel design, with constituent operations running across AIEs and connected by streams. Blue arrows are internal streams, green arrows are external streams between the AIEs and PL.}
  \label{fig:split_kernel}
\end{figure}

To address this we experimented with alternative kernel structures and, as illustrated by Figure \ref{fig:split_kernel}, split the code into multiple kernels each corresponding to a specific operation. By splitting apart the addition and multiplication, so each handles four of the eight calculations, we were able to increase the overall number of streams to six (two per kernel). There is a downside, as each individual kernel is now under utilised because it is now only undertaking four vectorized operations per cycle rather than eight, but this splitting results in six, rather than two, 128 bit streams connecting the PL to AIE kernel inputs. Consequently the HLS kernel running on the PL is able to stream the entirety of a grid cell's required data each cycle, reducing the initiation interval to one.

The performance of this approach is reported by the \emph{multi-kernel} row of Table \ref{tab:aie_performance}, and whilst this doubled performance compared to the initial version, it was still slower than the PL-only implementation. When undertaking profiling of our multi-kernel code using Vitis analyzer, we discovered that kernels were stalling on stream reads for over 60\% of the time. This is because, as described in Section \ref{sec:aie}, the physical streams between AI engines are 32 bits wide whereas per vectorized operation the kernel is generating 128 bits. Consequently the kernels were stalling waiting for the arrival of this data before operating upon it.

Connecting AIE kernels via cascade streams is an alternative approach and these, unlike the normal 32 bit streams, are 384 bits wide. We packed the 128 bit results into the cascade stream's \emph{accfloat} type, and streamed the entirety of the required data in one cycle. However, the limitation with cascade streams is that they physically connect between AIE cores by travelling in a horizontal manner, and when reaching the edge of a row connecting to the core above. Consequently their connection is inflexible, with each AIE core capable of only consuming cascade stream input from a single predefined neighbour and providing cascade stream output to its other neighbour. This is a problem for our multi-kernel design illustrated in Figure \ref{fig:split_kernel} as the \emph{subtraction-reduction} kernel requires inputs from two kernels, effectively requiring two cascade streams to feed into an AIE which is not possible on this architecture.

\begin{figure}[h]
  \centering
  \includegraphics[scale=0.8]{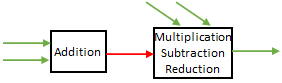}
  \caption{Cascade streaming approach, Red arrow is cascade stream, green arrows are external streams to/from the PL.}
  \label{fig:cascade_kernel}
\end{figure}

Therefore, to experiment whether cascade streams would improve performance, we adopted the design illustrated in Figure \ref{fig:cascade_kernel}, where one addition kernel undertakes all eight addition operations, and a separate kernel then undertakes the multiplication, subtraction, and reduction. The performance of this configuration is reported by \emph{cascade stream} in Table \ref{tab:aie_performance}, and the major reason for the poor performance is that the initiation interval on the PL increased to two as streams to the addition kernel require two writes per PL cycle as all eight pairs of operands are required by the single kernel. To address this we multiplexed the cascade streaming approach, with two separate copies on the AI engines such that, on average, over two clock cycles each AIE configuration receives its data. Performance of this approach is reported by the \emph{cascade multiplex} row in Table \ref{tab:aie_performance}, which improved performance but was still slower than that obtained by the non-AIE PL-only approach. Incidentally we also experimented with 4-way and 8-way multiplexing but this had no measurable improvement on performance.

\begin{figure}[h]
  \centering
  \includegraphics[width=\linewidth]{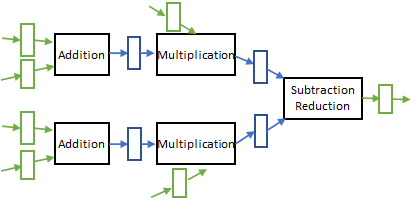}
  \caption{Multi-kernel windowing approach, coloured squares are the windows, blue connects kernels internally, and green arrows are external streams to/from the PL.}
  \label{fig:multi_kernel_windows}
\end{figure}

To this point we have explored connecting kernels and the PL via streams, however it is also possible to use windows which provide buffers. Importantly, an AIE can read up to 256 bits per cycle from memory compared to 32 bits from streams. Therefore we reverted to our multi-kernel design of Figure \ref{fig:split_kernel} and used windows instead of streams between the kernels as well as to drive input and output data between the PL. This is illustrated in Figure \ref{fig:multi_kernel_windows} and the performance is reported as \emph{multi-kernel windows} in Table \ref{tab:aie_performance}. It can be seen that the performance was extremely poor and this is because we were operating the windows on a grid cell by grid cell basis. This meant that there was no longer a pipelined loop within each kernel because between each grid cell the kernel was stopped and restarted by the AIE graph to fill and empty the windows as required by the AIE tooling.

We modified our windowing approach to work in chunks, where data for a number of grid cells is buffered into the windows and these operate ping-pong fashion where one copy is filled with data from the producer (either the PL or another AIE kernel) whilst the other window copy is being consumed, with these switched between outer iterations of the AIE graph. Consequently our addition, multiplication, and subtraction-reduction AIE kernels are concurrently processing different chunks of grid cells based upon the data available, effectively operating as a pipeline. An added complication was that because AIE kernels are operating out of sync, for example the multiplication kernels are one chunk behind their corresponding addition kernels, this stalled the PL. This was because when streaming data to the AIEs, writes to the multiplication streams are blocked waiting for the window to become free, but this waiting on the PL also blocks writes to the addition streams which are required to progress the addition AIE kernel which will unlock its multiplication kernel. The solution was to implement explicit ping-pong buffering on the PL for the multiplication streams, with a dataflow region working in sizes of \emph{chunk} which is concurrently filling a buffer with the current chunk's data and streaming out the previous chunks data to the AIE kernel.



Performance is reported by \emph{chunking windows} in Table \ref{tab:aie_performance}, where it can be seen that this approach has significantly increased performance on the AIEs, however it is still slightly slower than the PL-only. Based on profiling via Vitis analyzer we found that the subtraction and reduction kernel was taking around double the execution time of the other kernels and this imbalance of work was causing additional stalling. Consequently we modified the kernel to perform subtraction only and streamed back to the PL 4 floating point numbers which the PL then adds together. This is reported by the row \emph{reduction on host} in Table \ref{tab:aie_performance} which outperforms the PL-only. 

As described previously, in this multi-kernel approach each kernel is only working with vector sizes of four whereas the hardware is capable of undertaking eight single precision floating point operations per cycle. Working with windows, it was trivial to read two grid cells concurrently, placing the first in the lower portion of the vector and the second grid cell in the higher portion. Consequently this meant that vector operations were now running over eight operands, effectively processing two grid cells per AIE vectorized operation. This is reported by \emph{double vectorization} in Table \ref{tab:aie_performance} and resulted in a performance improvement, albeit modest as we are still limited to streaming data for only one grid cell between the PL and AIEs per cycle due to the maximum of port width of 128 bits.

\section{Multiple HLS compute units}
\label{sec:multiple}
In Section \ref{sec:aie_focus} we focused on a single PL HLS Compute Unit (CU). By decomposing across the advection problem's grid space, we can scale to multiple HLS CUs, all with a separate 3D part of the grid and working independently, connected to their own AIEs. Using our optimised AIE approach, which requires fifteen AIEs per HLS CU, we compared performance against other hardware options and Table \ref{tab:overall_performance} reports these results. 

The advection kernel running on the AI engines of the VCK5000 is reported by the row \emph{VCK5000 AIEs} in Table \ref{tab:overall_performance}. Whilst not documented directly, there are a maximum of 78 128-bit PLIO input streaming interfaces that only become apparent during compilation as we scaled. This is because AIE tile contains eight 32-bit AI Engine to AXI4-Stream channels \cite{aie_architecture} and there are 39 tiles. Consequently, there are a total of 312 32-bit channels connecting the AIEs to the PL, or a maximum of 78 128-bit channels as each of these is built using four 32-bit links. Incidentally AIEs accessing DRAM directly, without the PL, would also encounter this limitation as the data still needs to traverse these same physical links.

With six input streams per field, and three fields per CU, this results in a maximum of four HLS CUs. We are therefore using 60 AIEs in total, and up to four CUs the performance scales well. Consequently this hardware restriction is a major limitation because, if we were able to scale to a greater number of CUs, then performance would likely increase significantly. The importance of streaming an entire grid cell per cycle between the PL and AIEs was highlighted in Section \ref{sec:aie_focus}, and out of the two AIE kernel designs which enable this, multi-kernel and multiplexed cascade stream, the multi-kernel is preferable in this regard as it requires six input streams per field compared to eight for the multiplexed cascade stream. 

The Alveo U280 was configured with six HLS CUs, which is the maximum number that can fit due to limits on the number of ports in the Alveo shell. It can be seen that performance on the Alveo U280 is similar to that obtained on the VCK5000 using AI engines, even though there are only four CUs on the VCK5000. This is especially impressive considering that the U280 contains external HBM2 memory whereas the VCK5000 only has DDR4. We are able to fit eight CUs onto the PL-only VCK5000 configuration, which does not suffer from limitations on the number of ports due to the Versal containing a NoC which HLS kernels are connected to. The \emph{VCK5000 combined} result reports performance for a combination of the four AIE CUs with six PL-only CUs on the VCK5000, and this combined approach which leverages both the AIEs and PL for calculations delivers double the performance of the U280.

By comparison, the scheme running over the 24-core Cascade Lake Xeon Platinum CPU, which was threaded via OpenMP and compiled using GCC version 10.2 performs poorly compared with every other hardware technology. The V100 GPU version is implemented using OpenACC and version 20.9 of the Nvidia compiler, and this out-performs all other CPU and FPGA configurations, which is largely in agreement with \cite{brown2021accelerating}. Whilst the Versal has closed the gap with the GPU, it is unfortunate that AIE hardware restrictions ultimately limit the number of AIE CUs to four.

\begin{table}[htb]
 \begin{center}
  \begin{tabular}{|cc|}
    \hline
    \textbf{Description} & \textbf{Performance (GFLOPS)} \\
    \hline
    VCK5000 AIEs \textit{(4 CUs)} & 68.73 \\
    VCK5000 PL-only \textit{(8 CUs)} & 101.78 \\
    VCK5000 combined \textit{(4 and 6 CUs)} & 145.11 \\
    Alveo U280 \textit{(6 CUs)} & 72.32 \\
    24-core Xeon Platinum CPU & 23.52 \\
    V100 GPU & 227.89 \\
  \hline
\end{tabular}
  \caption{Compute performance of FPGA AIE and PL-only approaches compared to 24-core Cascade Lake CPU and Nvidia V100 GPU. All runs undertaken in single precision floating point using a problem size of 67 million grid points}
  \label{tab:overall_performance}
 \end{center}
\end{table}

\section{Conclusions and recommendations}
\label{sec:conclusions}
In this paper we have explored porting of the PW atmospheric advection scheme to the Versal, utilising the PL for tailoring memory accesses via a 3D shift-buffer and the AIEs for undertaking computation. Representative of a much wider class of stencil-based algorithms, which are popular in HPC workloads, we found that the major challenge was being able to most effectively interface the PL and AIEs to ensure data continually flows between the two. There are several possible approaches, and we have explored how hardware and tooling limitations drive specific choices and the performance impact of these. Ultimately, we found that the most effective approach was to use windows in a ping-pong fashion, working on chunks of data within the AIE kernels which rely on software pipelined loops and fully filled 8-way vectorization. Comparing against other hardware options, we found that a major limitation in obtaining performance was in the total number of streams between the PL and AIEs, which meant we were unable to scale beyond four HLS CUs. However four CUs using AIEs on the VCK5000 performed comparatively to six CUs on the Alveo U280 with the later benefiting from HBM2. The PL-only approach on the VCK5000 delivered impressive performance against the other FPGAs and CPU, which was largely due to being able to fit eight HLS CUs onto the PL, and when combining the AIEs and PL for compute we were able to deliver a significant improvement in performance compared to other FPGA approaches and the CPU. Therefore, AIEs aside, our PL-only experiments demonstrate that the Versal is a powerful architecture and improves on the Alveo.

From a development perspective there are many advantages in using the AIEs, and this will likely make the ACAP more accessible to software developers compared to traditional FPGAs. These include the overall compilation being much quicker, the ability to undertake much of the development exploration using simulation which itself is fast, no need to rebuild the PL if the interfaces between the PL and AIEs have not changed (which means bitstream regeneration takes around a minute), and the rich profiling tooling to provide insights where bottlenecks lie in the code. However it is crucial to match the workload to the architecture, and given the bandwidth between the PL and AIEs those kernels which have a higher FLOP to byte ratio than the stencil computation described in this paper will likely suit the AIEs much better. Therefore, an important lesson from this work is to focus primarily on those kernels that will not be limited by the current generation's PL to AIE interface, and algorithms with a high FLOP to byte ratio are likely where we will see the greatest benefit from this architecture.

Considering future enhancements to the Versal, in future AIE versions it would be beneficial if Xilinx were to make the physical streams between AIEs wider than 32 bits and increase the PL to AIE memory size from 128 to 256 bits, as well as supporting a larger number of PLIO streams. Increased flexibility around vector sizes would also be useful, for instance it is not possible to have a single precision floating point vector of numerous sizes including six, which required us to pad with empty values to eight, and this increased the amount of data transferred between PL and AIE. Considering the wider Vitis technology, within HLS it is not possible to create arrays of external AXI streams (e.g. of type \emph{qdma\_axis<128,0,0,0>}) and this resulted in messy code when experimenting with multiplexing. This is important because interfacing with AIEs will likely require a greater number of AXI streams compared to what is currently most common in HLS, and-so improved flexibility would be advantageous.

\section{Acknowledgements}
The authors would like to thank the ExCALIBUR H\&ES CGRA project who funded this work. We also acknowledge the ExCALIBUR H\&ES FPGA testbed and AMD Xilinx HACC program for access to compute resource used in this work, the later who also kindly provided comments and technical advice. For the purpose of open access, the author has applied a Creative Commons Attribution (CC BY) licence to any Author Accepted Manuscript version arising from this submission.

\bibliographystyle{ACM-Reference-Format}
\bibliography{sample-base}

\end{document}